\documentclass{elsart}
\usepackage{graphicx}
\usepackage{amssymb}
\usepackage{epsfig}
\usepackage{subfigure}
\usepackage{color}
\usepackage[english]{babel}
\usepackage{latexsym}
\usepackage{amsfonts}
\usepackage{amsmath}
\usepackage{multirow}
\usepackage{float}

\def\NPA#1#2#3{Nucl. Phys. A {\bf #1}\ (#3)\ #2}
\def\PLB#1#2#3{Phys. Lett. B {\bf #1}\ (#3)\ #2}
\def\PRC#1#2#3{Phys. Rev. C {\bf #1}\ (#3)\ #2}
\def\PRD#1#2#3{Phys. Rev. D {\bf #1}\ (#3)\ #2}

\def\ZPC#1#2#3{Z. Phys. C {\bf #1}\ (#3)\ #2}

\def\EPJC#1#2#3{Eur. Phys. J. C  {\bf #1}\ (#3)\ #2}

\def\RMP#1#2#3{Rev. Mod. Phys. {\bf #1}\ (#3)\ #2}

\def\RPP#1#2#3{Rep. on Progr. in Phys. {\bf #1}\ (#3)\ #2}

\def\be{\begin{equation}}
\def\ee{\end{equation}}

\newcommand{\ud}{\mathrm{d}}

\begin{document}
\begin{frontmatter}

\title{Interacting hadron resonance gas meets lattice QCD}

\author[gsi,mu]{A.~Andronic},
\author[gsi,emmi,tud,fias]{P.~Braun-Munzinger},
\author[hei]{J.~Stachel},
\author[hei]{M.~Winn}

\address[gsi]{GSI Helmholtzzentrum f\"ur Schwerionenforschung,
D-64291 Darmstadt, Germany}
\address[mu]{Institut f\" ur Kernphysik, 
Universit\"at M\" unster, D-48149 M\" unster, Germany}
\address[emmi]{ExtreMe Matter Institute, GSI, D-64291 Darmstadt, Germany}
\address[tud]{Technische Universit\" at Darmstadt, D-64289 Darmstadt, Germany}
\address[fias]{Frankfurt Institute for Advanced Studies, Goethe Universit\" at, D-60438 Frankfurt, Germany}
\address[hei]{Physikalisches Institut der Universit\"at Heidelberg,
D-69120 Heidelberg, Germany}

\begin{abstract}
  We present, in the framework of the interacting hadron resonance
  gas, an evaluation of thermodynamical quantities. The interaction is
  modelled via a correction for the finite size of the hadrons. We
  investigate the sensitivity of the model calculations on the radius
  of the hadrons, which is a parameter of the model.  Our calculations
  for thermodynamical quantities as energy and entropy densities and
  pressure are confronted with predictions using the lattice Quantum
  Chromodynamics (QCD) formalism.
\end{abstract}

\end{frontmatter}

\section{Introduction}
One of the major goals of ultrarelativistic nuclear collision studies
is to obtain information on the QCD phase diagram \cite{pbm_wambach}.
A promising approach is the investigation of hadron production.
Hadron yields measured in central heavy ion collisions from AGS up to
RHIC energies can be described very well
\cite{agssps,satz,heppe,cley,rhic,becgaz,aa05,aa08} within a
hadro-chemical equilibrium (also called hadron resonance gas or
statistical) model.  The main result of these investigations is that
the extracted temperature values rise rather sharply from low energies
on towards a center-of-mass energy per colliding nucleon pair
$\sqrt{s_{NN}}\simeq$10 GeV and reach, for higher collision energies,
constant values near $T$=160-165 MeV, while the baryochemical potential
$\mu_b$ decreases monotonically as a function of energy.  The Hagedorn 
limiting temperature \cite{hagedorn85} behavior suggests a connection to 
the phase boundary between the hadronic world and the deconfined phase.
It was, indeed, argued \cite{wetterich} that the quark-hadron
deconfinement phase transition drives the equilibration dynamically,
at least for SPS energies and above.  For lower energies, the
quarkyonic state of matter \cite{mclerran_pisarski} could complement
this picture. The conjecture of the triple point \cite{tri}
between hadronic, deconfined and quarkyonic matter was put forward in
this context. In a recent study \cite{floer} it is argued, however, 
that the chemical frezee-out region at large $\mu_b$ is not close to 
the phase boundary. 

Theoretical investigations of the QCD phase diagram are an important
priority for ongoing research.  While effective field theories need to
be employed to model QCD in the strongly interacting regime for finite
$\mu_b$ \cite{fuku-hat,fuku,herbst}, QCD calculations on the lattice are an
increasingly reliable approach for $\mu_b\simeq0$. Employing calculations 
of QCD on lattice all groups predict indeed a steep increase of thermodynamical
quantitites near a critical temperature for deconfinement, $T_c$.
Until recently, values for $T_c$ between 151 MeV \cite{aoki} and 192
MeV \cite{cheng1} were obtained.  New results on larger lattices and
with quark masses approaching the physical values \cite{aoki09,baz11}
lead to better agreement on the $T_c$ value in the range 155-160 MeV, 
while important details of lattice QCD calculations continue to be addressed. 
In this context, the hadron resonance gas (HRG) is used 
by lattice QCD groups as reference for their calculations in the hadronic 
sector \cite{huov09,bors2}. This follows earlier ideas of \cite{karsch03}
and involves modelling of quark mass dependence in the hadron resonance 
gas model to account for the finite lattice spacing 
\cite{huov09,bors2,karsch03}. 
Conversely, lattice data are used to constrain effective models based on 
hadronic resonances \cite{turko} to describe  hadronic matter near T$_c$.

The details of the hadron resonance gas (or statistical) model are
important too, as we have recently shown in \cite{aa08}, where we
demonstrated that the completeness of the hadron spectrum involved in
calculations is important for a precise description of data in
nucleus-nucleus collisions.  The aim of this paper is to confront our
HRG model calculations with lattice QCD predictions for
thermodynamical observables in the hadronic sector.  In particular, we
investigate the effect of the excluded volume correction employed in
the model to approximate a short-range repulsive hadron-hadron
interaction. Such excluded volume corrections were first introduced in
\cite{hagedorn80}, albeit not yet in a thermodynamically consistent
way. A thermodynamically consistent approach was first developed in
\cite{exv} and will be the basis for our investigations.

\section{Model description}

We restrict ourselves here to the basic features and essential results
of the statistical model approach. A complete survey of the
assumptions and results, as well as of the relevant references, is
available in ref.~\cite{review}.

The basic quantity required to compute the thermal composition of
hadron yields and the thermodynamical quantities is the partition
function $Z(T,V)$. In the grand canonical (GC) ensemble, the partition
function for a particle species $i$ in the limit of large volume takes
the following form ($k=\hbar=c=1$): 
\be \ln Z_i^{id.gas} ={{Vg_i}\over
  {2\pi^2}}\int_0^\infty \pm p^2\ud p \ln [1\pm \exp
  (-(E_i-\mu_i)/T)],
\label{eq:part}
\ee
from which the particle density $n_i$, the partial pressure $P_i$, the energy 
density $\varepsilon_i$ and the entropy density $s_i$ are then calculated 
according to:
\begin{align}
n_i^{id.gas}(T, \mu_i)&=N_i/V=\frac{T}{V}\left(\frac{\partial\ln Z_i^{id.gas}}
{\partial\mu}\right)_{V,T}=\frac{g_i}{2\pi^2} \int_0^\infty \frac{p^2 \ud p}
{\exp[(E_i-\mu_i)/T] \pm 1} \label{eq: density}\\
P_i^{id.gas}(T, \mu_i)&= \frac{T}{V} \ln Z_i^{id.gas} = \pm
\frac{g_{i} T} 
{2 \pi^{2}} \int^{\infty}_{0} p^2 \ud p  \ln \left(1\pm \exp[-(E_i-\mu_i)/T]\right) 
\label{eq: pressure}\end{align}
\begin{align}
\varepsilon_i^{id.gas}(T, \mu_i)&= E_i/V = - \frac{1}{V} \left( \frac{\partial 
\ln Z_i^{id.gas}}{\partial(1/T)} \right)_{\mu/T} =
\frac{g_{i}} {2 \pi^{2}} \int^{\infty}_{0} \frac{ p^2 \ud p }
{\exp[(E_i-\mu_i)/T]\pm 1}  E_i \label{eq: energy density}\\
s_i^{id.gas}(T, \mu_i) &= S_i/V = \frac{1}{V} \left(\frac{\partial 
(T \ln Z_i^{id.gas})}{\partial T}\right)_{V,\mu} = \nonumber \\
&\pm \frac{g_{i}} {2 \pi^{2}} \int^{\infty}_{0}p^2 \ud p  
\left( \ln \left(1\pm \exp[-(E_i-\mu_i)/T]\right) \pm 
\frac{E_i-\mu_i}{T(\exp[(E_i-\mu_i)/T] \pm 1)}\right),\label{eq: entropy density}
\end{align}
where $g_i=(2J_i+1)$ denotes the spin degeneracy factor, $T$ is the
temperature and $E_i =\sqrt {p^2+m_i^2}$ is the total energy.  The
\mbox{(+)} sign corresponds to fermions and \mbox{(--)} corresponds to
bosons.  For the hadron species $i$ of baryon number $B_i$, third
component of the isospin $I_{3i}$, strangeness $S_i$, and charm $C_i$,
the chemical potential is $\mu_i = \mu_b B_i+\mu_{I_3} I_{3i}+\mu_S
S_i+\mu_C C_i$.  The chemical potentials related to baryon number
($\mu_b$), isospin ($\mu_{I_3}$), strangeness ($\mu_C$) and charm
($\mu_C$) ensure the (on average) conservation, in the collision, of
the respective quantum numbers: 
i) isospin: $V_{cons} \sum_i n_i I_{3i} = I_{3}^{tot}$, with $V_{cons}=N_B/\sum_i n_i B_i$;
ii) strangeness: $\sum_i n_i S_i = 0$;
iii) charm: $\sum_i n_i C_i = 0$. The (net) baryon number $N_B$ and the total
  isospin $I_{3}^{tot}$ of the system are input values which need to
  be specified according to the colliding nuclei and rapidity interval
  studied.  Taking into account the conservation laws (i-iii), the
  freeze-out temperature $T$, the baryochemical potential $\mu_b$ and 
 the fireball volume at chemical freeze-out $V$ are the parameters of 
the model, which are obtained from fits to experimentally measured hadron 
yields.  
\\ 
The following hadrons (number of species, not counting $g_i$) are
included in the calculations: i) mesons: non-strange (123), strange (32), 
charm (40), bottom (28); ii) baryons: non-strange (48), strange (48), 
charm (32), bottom (14). The corresponding anti-particles are of course 
also included.  Their characteristics, including a rather complete set of 
decay channels (all strong and electromagnetic decays), are implemented
according to the 2008 PDG compilation\footnote{The 2010 PDG
  compilation contains updates in the hadron mass
  spectrum, but these are expected to have a minor influence on our results.} 
\cite{pdg}, with hadron masses reaching 3 GeV.  We use vacuum masses for 
all hadrons.

Usually, whenever thermal fits are performed, the finite widths of 
resonances are taken into account in the density calculation by an 
additional integration, over the particle mass, with a Breit-Wigner 
distribution as a weight \cite{aa05}.
For the  range of temperatures investigated in this work the effect of 
the finite resonance widths is small and to save computing time we have
not employed the additional integration 
in the present calculations except where stated otherwise.

\section{Interactions in the hadron gas model}

When comparing thermodynamical quantities computed within the
framework of the hadron resonance gas model with results obtained
using lattice QCD methods one has to decide how to incorporate
interactions among the hadrons. One approach is to use results
obtained by the authors of \cite{beth-uhlenbeck,dashen1,dashen2} where
two-body collisions are taken into account through scattering phase
shifts. Here the interaction measure (the 2nd virial coefficient) is
related to the derivative of the phase shifts with respect to energy. 
To compute the thermodynamics of the interacting hadron
resonance gas in this way one would need knowledge of the energy
dependence of all phase shifts. At first glance this seems quite
impractical. An interesting result was, in this context, obtained in
\cite{prakash}. These authors show by explicit construction that, for
simple systems such as gases of pions, pions and nucleons, and pions,
kaons, and nucleons, the equation of state of the interacting system
is obtained by adding the relevant resonances, the $\rho$ and
$f^0$(980) mesons, the $\Delta$ baryon, the $K^*$ meson, to the list
of particles and by computing the partition function of the enlarged
gas assuming no interactions.

This interesting result has led some authors \cite{huov09,bors2} 
to argue that the thermodynamics of the interacting hadron resonance gas 
is well approximated, via the Dashen, Ma and Bernstein theorem
\cite{dashen1,dashen2}, by that of the non-interacting case, provided
that {\it all} states (resonances) are included in the partition
function.  It is one of the goals of this paper to address the
accuracy of this approximation within the framework of our interacting
HRG model.  Even at the formal level, there are points to be
considered.  First, the $\eta, \omega, \eta', \phi$ and a$_0$ mesons
cannot be treated like this \cite{prakash}. Second, the baryon-baryon
interaction is largely repulsive, with no known resonance
structure,\footnote{We neglect here the deuteron in the $^3$S$_1$
state. In the baryon-antibaryon system there is likely no short-range 
repulsion and this leads to a small correction which is discussed below.} 
see, e.g. fig. 8 of \cite{prakash}. 
More importantly, the approach of \cite{dashen1,dashen2} is, as also 
discussed there, a low density approach, relevant for dilute systems. 
At temperatures near
$T_c$, the temperature of the phase boundary between hadron gas and
quark-gluon plasma, the hadron resonance gas is not dilute anymore. As
will be shown below, overall densities exceed 0.5 fm$^{-3}$ and total
baryon densities are close to normal nuclear matter densities of
0.15 fm$^{-3}$. This implies that for the whole range of baryon chemical
potentials considered here the baryon densities near $T_c$ are close
to or exceed the critical value worked out in \cite{prakash} above
which the virial expansion breaks down.  In this environment also the
concept of asymptotic states needed for the S-matrix approach of
\cite{beth-uhlenbeck,dashen1,dashen2} is ill defined.

We therefore explore in the following, in addition to the 'free'
hadron resonance gas, also the thermodynamic properties of a hadron
resonance gas in which short-distance repulsion is explicitely taken
into account using the thermodynamically consistent excluded volume
approach developed in \cite{exv}. In essence this amounts to a
Van-der-Waals construction.
This is implemented according to  \cite{exv,exv2} in an iterative procedure
for the total pressure as:
\be 
P(T,\mu_1,...,\mu_m)= P^{id.gas}(T,\hat{\mu}_1,...,\hat{\mu}_m)
\ee
where $P^{id.gas}=\sum_i  P^{id.gas}_i(T,\hat{\mu}_i)$  and for each particle $i$
the chemical potential at a given iteration is recalculated as:
\be
\hat{\mu_i} = \mu_i - V_{eigen, i} P(T,\mu_1,...,\mu_m). \ee 
This approach yields the following formulae for the particle densities 
$n_i$, the total energy density $\varepsilon$ and
the total entropy density $s$ expressed as a function of the respective 
quantities in the ideal gas, which are given in 
(\ref{eq: density}-\ref{eq: entropy density}):
\begin{align}
n_i=n_i(T,\mu_1,...,\mu_m) &= \left ( \frac{\partial P}{\partial 
\mu_i}\right)_T = \frac{n_i^{id.gas}(T, \hat{\mu_i})}{1+\sum_k V_{eigen,k} 
n_k^{id.gas}(T, \hat{\mu_k})} \label{eq: Teilchendichteexcl}\\
s=s(T,\mu_1,...,\mu_m) &=\left( \frac{\partial P}{\partial T}
\right)_{\mu_1,...,\mu_m} = \frac {\sum_i s_i^{id.gas}(T,\hat{\mu}_i)}{1+ 
\sum_k V_{eigen,k} n_k^{id.gas}(T, \hat{\mu_k})} \label{eq: Entropieexcl}\\
\varepsilon=\varepsilon(T,\mu_1,...,\mu_m) & =  
\frac{\sum_i \varepsilon_i^{id.gas}(T,\hat{\mu_i})}
{1+\sum_k V_{eigen,k} n_k^{id.gas}(T,\hat{\mu_k}) } \label{eq: Energieexcl}
\end{align}
where $V_{eigen, i}=4\cdot 4\pi R_i^3/3$ is the eigenvolume of a hadron\footnote{
Consider a particle with radius $R$ in the hard sphere model. Then no other 
particle can come closer than a distance 2$R$. Per pair the excluded volume 
is $4\pi (2R)^3/3$, leading to $V_{eigen}=4\cdot 4\pi R^3/3$ for the particle.} 
with radius $R_i$.
We checked numerically that thermodynamical consistency, expressed by 
$\varepsilon=Ts-P+\sum_i \mu_i n_i$, 
is well fulfilled by our calculations, explicitly confirming the consistency
of the procedure \cite{exv} used for the excluded volume correction.

For the radius parameter $R_i$, governing the excluded volume
calculation, we follow the earlier arguments in \cite{heppe}. There it
was argued that, for baryons, the radius is given by the hard-core
repulsive interaction as extracted from nucleon-nucleon scattering
\cite{bohr}, giving a radius of about 0.3 fm. Values for other baryons
should be similar. For mesons, in the absence of detailed information
on their interactions at short distance, we assign the same radius
value, based on the similarity of the meson charge radii compared to
baryons and on the energy dependence of the pion-nucleon phase 
shifts \cite{roper64}.
For illustration, we have included the case of $R$=0 for mesons, although
we believe that the physical case is for mesons with repulsive core radius 
$R$ comparable to baryons.

\begin{figure}[hbt]
\centering\includegraphics[width=.6\textwidth]{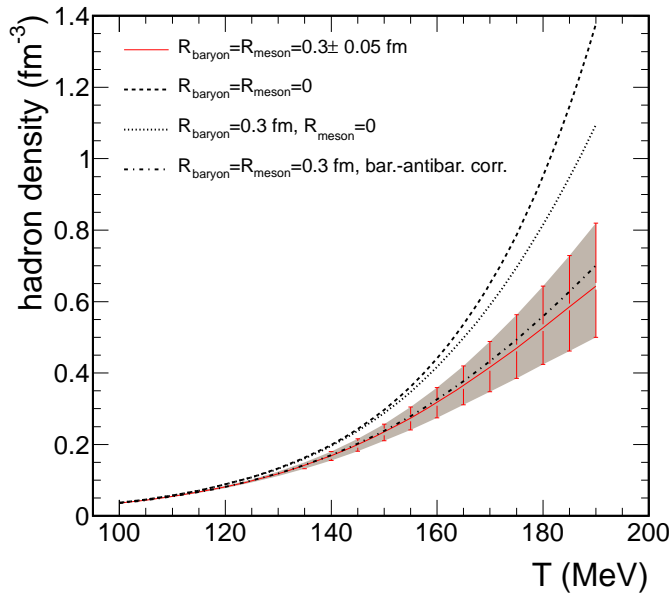}
\caption{Total hadron densitites as a function of temperature, calculated for
the hadron resonance gas model at $\mu_b$=0, without (dashed line) and with 
(band for R=0.3$\pm$0.05 fm) the excluded volume correction. The case of 
$R_{\mathrm{meson}}$=0 is shown with the dotted line, while the dot-dashed line 
denotes the effect of the baryon-antibaryon annihilation correction.}
\label{fig1}
\end{figure}

For the baryon-antibaryon system there are likely no short-range repulsive 
interactions, because of annihilation processes, which are by construction
included in the hadron resonance gas at equilibrium.
We have modelled the absence of short-range repulsion in a schematic way,
introducing a correction factor $k_{anni}$ for the (anti)baryon eigenvolume 
based on the expression:
\begin{equation}
k_{anni} = 1 - \frac{2n_{\text{baryons}}n_{\text{antibaryons}}}{(n_{\text{baryons}}+n_{\text{antibaryons}})^2},
\end{equation}
where $n_{\text{baryons}}$ is the density of baryons and  $n_{\text{antibaryons}}$ 
the density of antibaryons. As shown by the results presented in the 
following, the absence of short-range repulsion in the baryon-antibaryon
system leads to only a small correction, since mesons dominate 
at small $\mu_b$ and since there are very few antibaryons at large $\mu_b$.

For the rest of the paper we show results of our calculations for radii 
in the range of 0.3$\pm$0.05 fm, along with the cases discussed above and 
contrast those with the case of no interaction ($R_i$=0).
The value of 0.3 fm, common for mesons and baryons, was also used whenever 
we performed thermal fits to hadron abundancies \cite{heppe,rhic,aa05,aa08}.
In the description of hadron yields with the statistical model, the excluded 
volume correction leads to a larger volume parameter, while the fit 
temperature and baryochemical potential are unchanged compared to the case 
of fitting hadron ratios \cite{aa05}, for the case of identical $R_i$ for
mesons and baryons.
The implication of a pion radius different from all other hadrons 
for the description of data has been studied by Yen et al. \cite{exv2}.

In Fig.~\ref{fig1} we present the temperature dependence of the hadron
densities calculated with our model, with and without interactions
modelled via excluded-volume corrections, as well as with the cases of 
$R_{\mathrm{meson}}$=0 and of absence of short-range interactions for 
baryon-antibaryon pairs.  
This illustrates our remarks above, namely that, while at low temperatures 
(low densities) there is no difference between the case of excluded 
volume correction and the case of free hadron resonance gas, the difference 
becomes appreciable for $T$ above 130-140 MeV, which is significantly below
the value for the critical (crossover) temperature $T_c\simeq$160 MeV.
Near $T_c$, the hadron resonance gas becomes manifestly dense, with the mean 
distance between hadrons getting significantly smaller than twice the hadron 
radius. All approximations appropriate for the dilute gas, discussed above,
break down and the non-interacting hadron gas is not a suitable approach 
anymore.

\begin{figure}[hbt]
\centering\includegraphics[width=.6\textwidth]{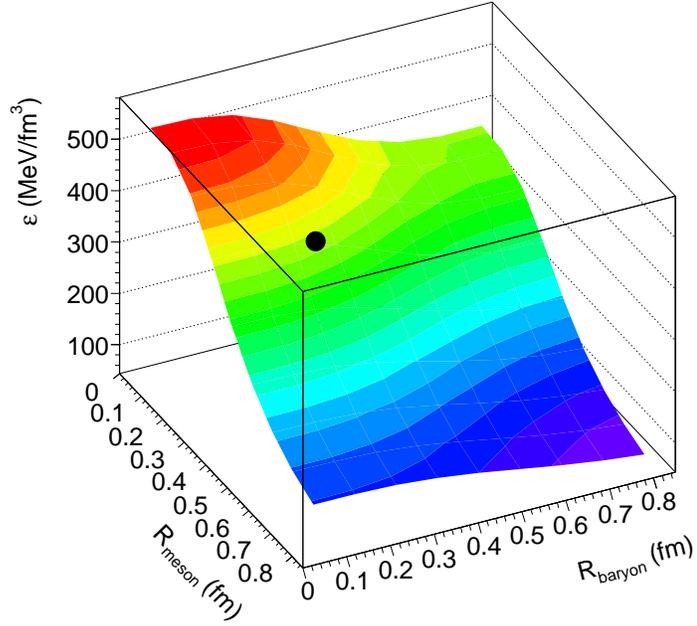}
\caption{The energy density of a hadron gas as a function of the radius 
(for the eigenvolume calculation) for mesons and baryons, at a
temperature value of 164 MeV and $\mu_b$=0.8 MeV. The dot indicates
the radius value of 0.3 fm which we employ as default.}
\label{fig2}
\end{figure}

We illustrate, for the energy density, $\varepsilon$, in Fig.~\ref{fig2} 
the sensitivity of the calculations on the radii of the eigenvolume for 
mesons and baryons, $R_{\mathrm{meson}}$ and $R_{\mathrm{baryon}}$, respectively.
The calculations have been performed for $T$=~164~MeV, corresponding to the
limiting temperature reached in heavy-ion collisions \cite{aa08} and for 
$\mu_b$=0.8~MeV, the value expected for the LHC energy according to 
ref.~\cite{aa08}; calculations for $\mu_b$=0 lead to identical results.
We observe a strong influence of the excluded volume correction on the
energy density and this is the case for all other thermodynamical quantities.
Due to the larger abundance of mesons (and in particular of pions) 
in the hadron gas at these values of $T$ and $\mu_b$, the sensitivity 
on $R_{mathrm{meson}}$ is more pronounced.

\begin{figure}[hbt]
\centering\includegraphics[width=.6\textwidth]{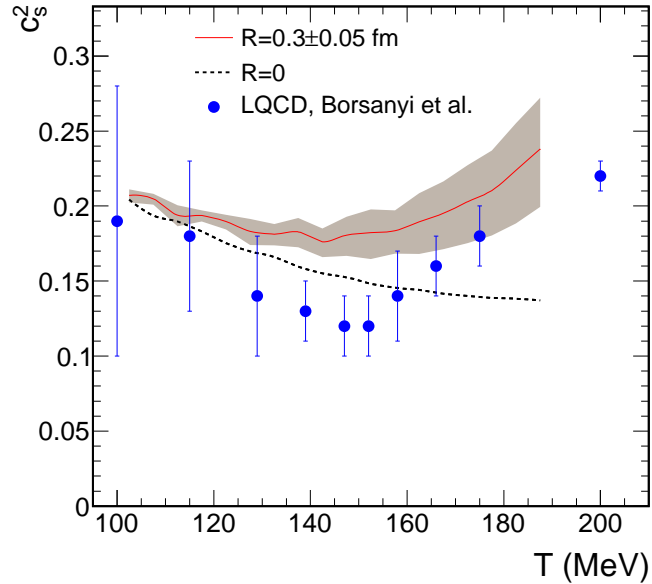}
\caption{The adiabatic speed of sound (in units of velocity of light, $c$=1) 
squared as a function of temperature, calculated for the hadron gas model, 
with and without excluded-volume corrections. Our calculations are compared 
to lattice QCD calculations of Bors\' anyi et al. \cite{bors}.}
\label{figx}
\end{figure}

As pointed out earlier \cite{exv,prakash}, a possible problem of the hadron gas
model with excluded volume corrections is acausal behavior (speed of sound
larger than velocity of light). As we demonstrate in Fig.~\ref{figx}, our model
is not plagued by such a behavior (for the case $\mu_b\simeq$0 considered here).
The adiabatic speed of sound, $c_s$, is calculated as:
\be
c_s^2=\left(\frac{\ud\ln s}{\ud\ln T}\right)^{-1}.
\ee
It exhibits a shallow minimum as a function of $T$ for the case of excluded 
volume corrections.
Our calculations are compared to lattice QCD calculations \cite{bors}.
In general, a good agreement between our and the lattice result is observed.
We note that our calculations predict a shallow minimum for $T$ around 
140-150 MeV, while the lattice values indicate a more pronounced dip and
exhibit a speed of sound value smaller than the hadron resonance gas
with interactions.
It would be interesting to see if the corresponding low temperature part
of the equation of state (EoS) would lead to changes in hydrodynamic 
calculations, where generally the EoS shown in Fig. 6 or ref. \cite{huov09} 
is used.

\section{Hadron resonance gas and lattice QCD results}

\begin{figure}[hbt]
\centering\includegraphics[width=.56\textwidth]{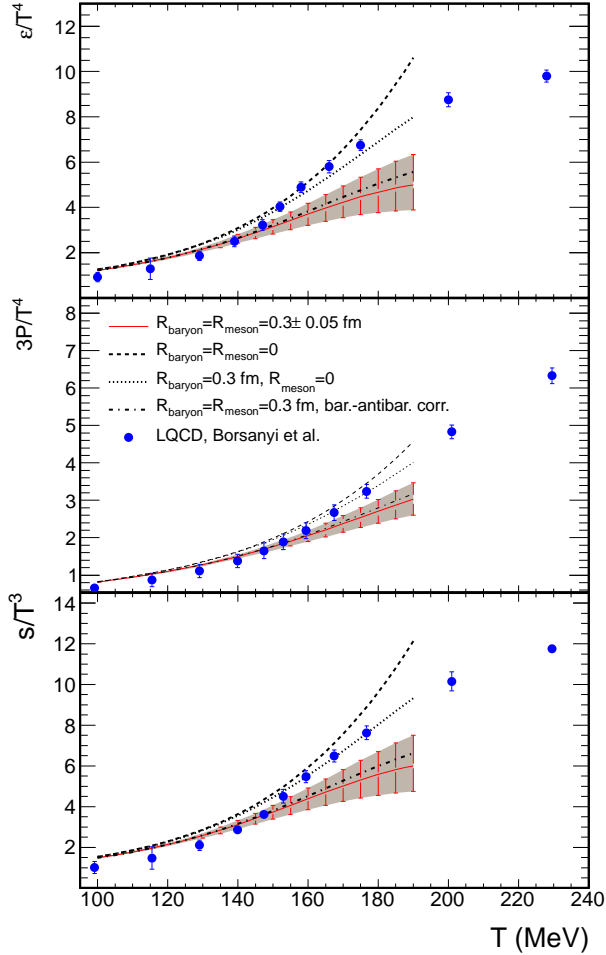}
\caption{Temperature dependence of thermodynamical quantities. The calculations
with the hadron gas model are shown without (dashed line) and with (band for 
R=0.3$\pm$0.05 fm) the excluded volume correction. The case of 
$R_{\mathrm{meson}}$=0 is shown with the dotted line, while the dot-dashed line 
denotes the effect of the baryon-antibaryon annihilation correction.
They are compared to LQCD results of Bors\' anyi et al. \cite{bors}.}
\label{fig3}
\end{figure}

In the following we compute, in the HRG model, thermodynamical
quantities with and without excluded volume corrections and compare
the results to predictions from lattice QCD.  In Fig.~\ref{fig3} we
show the temperature dependence of energy density, pressure, and
entropy density, each normalized to appropriate powers of the
temperature.

The case without interactions (no excluded volume correction) has the
expected strong dependence on temperature. As noted early on by
Hagedorn \cite{hagedorn85}, a limiting temperature, also called
``Hagedorn temperature'', of $T_H\simeq 200$ MeV arises for
calculations of thermodynamical quantities within the HRG model if one
assumes a hadron mass spectrum which increases exponentially with
particle mass. Such exponential behavior is consistent with the
present knowledge of hadron resonances \cite{pdg} up to 2.0-2.5 GeV in
mass. At this temperature all thermodynamical quantities for the HRG
without excluded volume corrections diverge. We note in passing that
all thermal model calculations without excluded volume corrections
become meaningless for temperature values close to $T_H$. In the
course of investigations reported in \cite{aa08} we realized that this
implies a practical limitation to temperatures below 175 MeV as all
calculations for higher temperatures become very sensitive to details
of the mass spectrum for masses larger than 3 GeV.

For the case of calculations employing finite hadron volume
corrections the Hagedorn infinities are tamed. This was already noted
by Hagedorn \cite{hagedorn85} who was the first to introduce excluded
volume corrections \cite{hagedorn80}. 
Our findings substantiate this and imply that the Hagedorn limiting 
temperature is an artifact of the usage of the free hadron resonance gas 
description at temperatures where the implicit approximations for dilute 
systems are manifestly inappropriate.

\begin{figure}[hbt]
\centering\includegraphics[width=.65\textwidth]{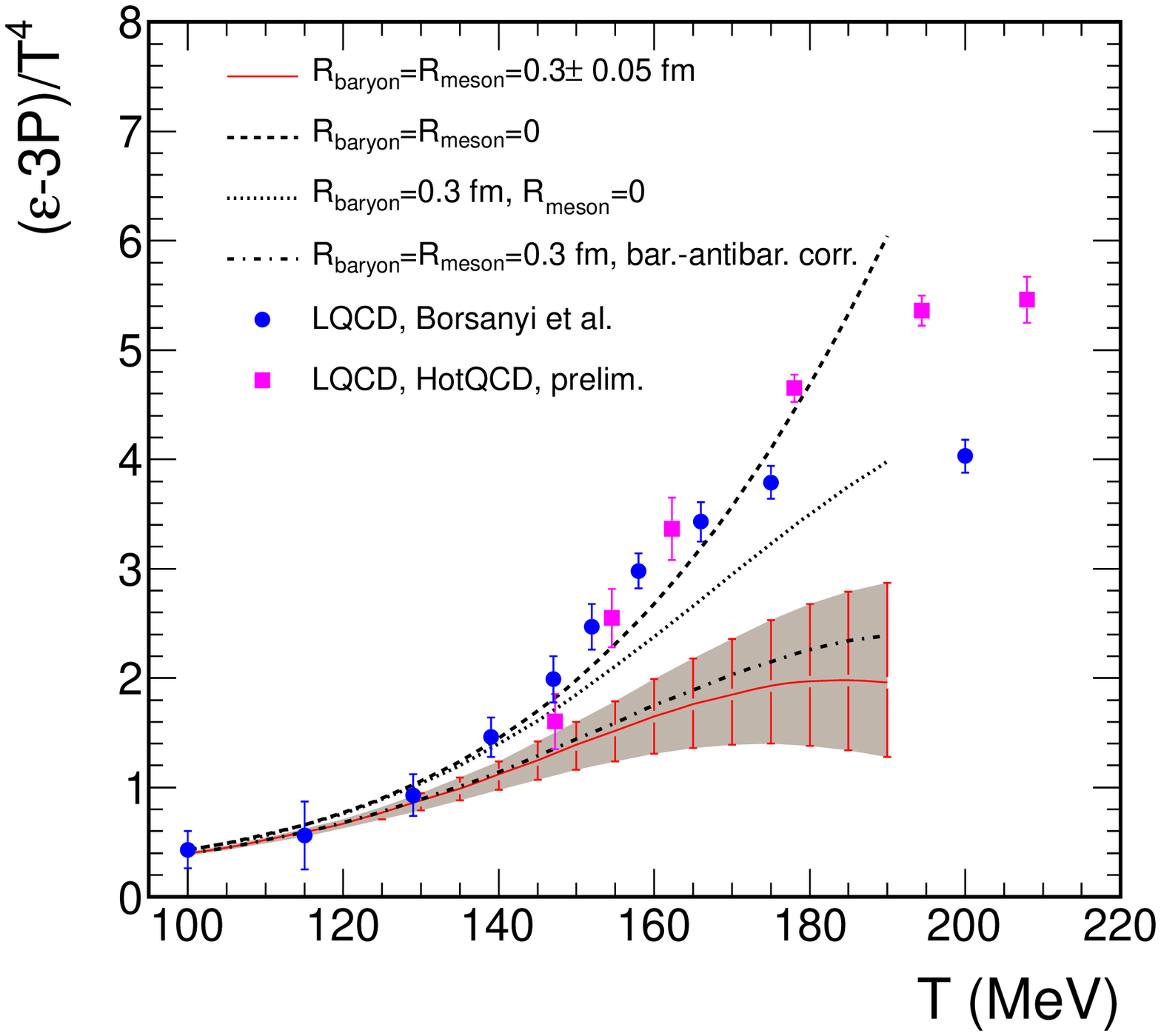}
\caption{Temperature dependence of the trace anomaly. The calculations
within the hadron gas model (lines, as in Fig.~\ref{fig3}) are compared 
to LQCD calculations of Bors\' anyi et al. \cite{bors} and HotQCD 
collaboration \cite{hotqcd} (preliminary results).} 
\label{fig3b}
\end{figure}

For temperatures below 120 MeV the HRG model results with and without 
excluded volume correction almost coincide, see Fig.~\ref{fig3}. 
For larger temperatures, the HRG with interactions yields, in our view, 
a realistic description of the hadronic phase. 
Therefore, in the confined regime, lattice QCD calculations of thermodynamical
variables should give results in agreement with the interacting HRG. 
The expectation is that, as soon as effects of deconfinement become 
important in the lattice QCD results, they should increasingly exceed 
the HRG values. In Fig.~\ref{fig3} the most recent predictions of lattice
QCD are compared to the HRG results.
Indeed, below $T$=150 MeV good agreement between results of
lattice QCD \cite{bors} and the interacting HRG is found.
On the other hand, effects of the onset of deconfinement \cite{aoki09,baz11}
are apparent for $T$ in excess of 150 MeV. 

In Fig.~\ref{fig3b} we compare our calculations for the trace anomaly 
$\varepsilon-3P$ (normalized to $T^4$) to LQCD results \cite{bors,hotqcd}.
We see an agreement between LQCD data and our calculations for the 
interacting hadron gas only up to $T$=140 MeV for the data of Bors\' anyi 
et al. \cite{bors}.
For the (preliminary) data of the HotQCD collaboration \cite{hotqcd} we have
used for illustration the set for hisq action with $N_t$=6, but we note 
that other available sets are in agreement with those within the 
errors \cite{hotqcd}.

\begin{figure}[hbt]
\centering\includegraphics[width=.64\textwidth]{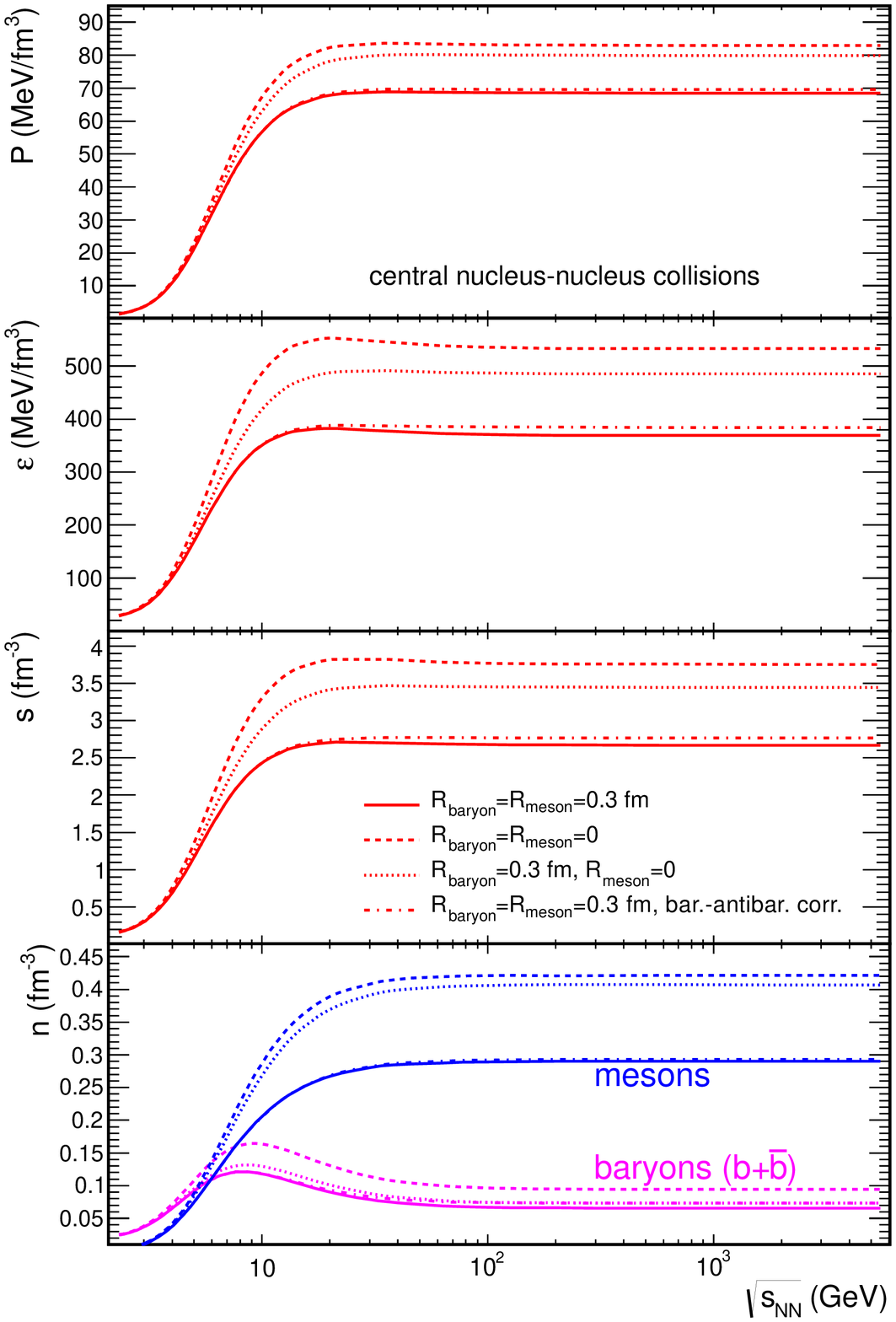}
\caption{Energy dependence of energy density, pressure, entropy density
and baryon and meson densities at chemical freeze-out in central
nucleus-nucleus collisions. The full lines are for the excluded volume 
corrections, the dashed line without, while the case of $R_{mathrm{meson}}$=0 
is shown with the dotted line and the dot-dashed line denotes
the effect of the baryon-antibaryon annihilation correction.}
\label{fig4}
\end{figure}

We turn now to the energy dependence of the thermodynamical quantities
at chemical freeze-out in central nucleus-nucleus collisions
\cite{aa08}. The degree of stopping of the colliding nuclei, which is
energy dependent, brings some uncertainty in the choice of $N_B$ and
$I_{3}^{tot}$. As we study central collisions of heavy nuclei (Au or
Pb) and focus on data at midrapidity, we have chosen $N_B=400 \cdot
\mu_b/938\ \text{MeV}$ and $I_{3}^{tot}= -40 \cdot
\mu_b/938\ \text{MeV}$, reflecting that $\mu_b$ traces stopping of the two
colliding nuclei. The sensitivity of the thermodynamical quantities on 
$N_B$ and $I_{3}^{tot}$ is, however, small. 

In Fig.~\ref{fig4} we show the thermodynamical quantities as a
function of collision energy for chemical freeze-out in central
nucleus-nucleus collisions. 
The trends seen in Fig.~\ref{fig4} reflect primarily the sharp increase 
of the temperature at chemical freeze-out (determined from fits of
experimental data up to $\sqrt{s_{NN}}$=200 GeV \cite{aa08}) followed by a
saturation above $\sqrt{s_{NN}}\simeq$10 GeV. The characteristic
energy dependence of the baryon density, exhibiting a maximum around 
8 GeV, is determined by the increase of $T$ combined with the strong
decrease of $\mu_b$ with energy, as discussed in \cite{aa08}.
The effect of interactions leads to up to 30\% reduction of the 
thermodynamical quantities at chemical freeze-out in nucleus-nucleus 
collisions.

\section{Summary}
We have presented an evaluation of thermodynamical quantities in the
framework of the interacting hadron gas model, incorporating all known
hadrons with masses reaching 3 GeV. A Van-der-Waals-type interaction
is modelled via an excluded volume correction.
Thermodynamic consistency is ensured by construction and the model
exhibits proper causal behavior.

The resulting values for the thermodynamical quantities increase,
already for temperatures of 130-150 MeV, i.e. significantly below $T_c$,
much less steeply than in case of a free hadron resonance gas.
Near $T_c$ the free hadron resonance gas calculations
already show signs of the Hagedorn divergence. Comparisons of lattice
QCD results with the free hadron resonance gas in this temperature
regime are therefore in our view problematic. Our results imply the
need to consider the hadron resonance gas with interactions, 
beyond the usual implementations based on the Dashen, Ma and Bernstein 
theorem.

On the other hand, lattice QCD simulations start to be precise enough to
reproduce the complete hadron gas at low temperatures. The apparent
rise of the lattice QCD results above the HRG results is a clear
indication of the onset of deconfinement not contained in the
latter.
In our view, the lattice results show genuine quark and gluon degrees of 
freedom in the vicinity of the (crossover) transition. In this temperature 
range, the lattice results produce thermodynamical quantities well above our
predictions for the interacting hadron resonance gas.
Our findings also imply that the Hagedorn limiting temperature is 
an artifact of the usage of the free hadron resonance gas description
at temperatures where the gas becomes manifestly dense.

\section*{Acknowledgments}
We thank Z. Fodor and F. Karsch for sending us the numerical values of 
their lattice QCD calculations and K. Redlich for discussions and reading
of the manuscript.

\end{document}